\documentclass[aps,pra,reprint, amsmath, amssymb,superscriptaddress]{revtex4-2}

\usepackage{dsfont}
\usepackage{graphicx} 
\usepackage{graphics}
\usepackage{verbatim}
\usepackage{mathtools}
\usepackage{dcolumn}    
\usepackage{bm} 
\usepackage{graphicx}
\usepackage{amsmath}    
\usepackage{latexsym}
\usepackage{amsfonts}   
\usepackage{amssymb}
\usepackage{array}      
\usepackage{epsfig}
\usepackage{txfonts}
\usepackage{xcolor}
\usepackage[normalem]{ulem}
\usepackage[colorlinks=true,linkcolor=blue,urlcolor=blue,citecolor=blue,pdfusetitle]{hyperref}
\usepackage{hyperref}
\usepackage{framed}
\newcommand{\Tr}{\mbox{Tr}}
\newcommand{\ket}[1]{\left|#1\right\rangle}
\newcommand{\bra}[1]{\langle#1|}
\newcommand{\braket}[2]{\langle#1|#2\rangle}
\newcommand{\ketbra}[2]{|#1\rangle\langle#2|}

\newcommand{\avg}[1]{\langle#1\rangle}

\begin{document}

\title{Metrological approach to the emergence of classical objectivity}

\author{Anthony Kiely}
\address{School of Physics, University College Dublin, Belfield Dublin 4, Ireland}
\affiliation{Centre for Quantum Engineering, Science, and Technology,
University College Dublin, Belfield, Dublin 4, Ireland}
\affiliation{School of Physics, University College Cork, College Road, Cork, Ireland}
\author{Diana A. Chisholm}
\affiliation{Universit\`a degli Studi di Palermo, Dipartimento di Fisica e Chimica - Emilio Segr\`e, via Archirafi 36, I-90123 Palermo, Italy}
\affiliation{School of Physics, University College Dublin, Belfield Dublin 4, Ireland}
\author{Akram Touil}
\address{Theoretical Division, Los Alamos National Laboratory, Los Alamos, New Mexico 87545}
\author{Sebastian Deffner}
\affiliation{Department of Physics, University of Maryland, Baltimore County, Baltimore, MD 21250, USA}
\affiliation{Quantum Science Institute, University of Maryland, Baltimore County, Baltimore, MD 21250, USA}
\affiliation{National Quantum Laboratory, College Park, MD 20740, USA}
\author{Gabriel Landi}
\address{Department of Physics and Astronomy, University of Rochester, Rochester, New York 14627, USA}
\address{University of Rochester Center for Coherence and Quantum Science, Rochester, New York 14627, USA}
\author{Steve Campbell}
\address{School of Physics, University College Dublin, Belfield Dublin 4, Ireland}
\affiliation{Centre for Quantum Engineering, Science, and Technology,
University College Dublin, Belfield, Dublin 4, Ireland}

\begin{abstract}
We present a precise characterization of the onset of classicality that combines the formalism of quantum Darwinism with the tools from quantum metrology. We show that the quantum Fisher information provides a useful metric for assessing the rate at which classical objectivity emerges. Furthermore, our formalism allows us to explore how the choice of measurement impacts the precision with which an observer can determine the state of the system. For a paradigmatic example of the spin-star model, we demonstrate that optimal measurements lead to the emergence of classicality at an exponential rate. Although other measurements necessarily lead to slower emergence, we importantly show that suboptimal measurements can still saturate the Cram\'{e}r-Rao bound. By recasting emergent classicality as an information acquisition protocol, our framework provides a precise operational description of quantum Darwinism.
\end{abstract}

\maketitle

\section{Introduction}

Quantum Darwinism (QD) provides a framework to explain how quantum systems, which are inevitably open to their environment, can exhibit objective properties, a signature trait of classical systems. As such QD is considered a first important step in unveiling the quantum-to-classical transition~\cite{Zurek2000AP,Ollivier2004PRL,Ollivier2005PRA,Zurek2009NP,zurek2022entropy,Korbicz2021roadstoobjectivity,Zurek_2025}. To date, many aspects of QD have been explored, including the effects of non-Markovianity~\cite{Milazzo2019, TaysaPRA}, noncommutativity~\cite{Ryan2022, chisholm2025emergence}, contextuality~\cite{Baldijao2021}, the role of consensus~\cite{chisholm2023meaning, touil2025consensus}, and the compatibility of measurements \cite{Doucet2025PRA}. A variety of models have been shown to exhibit the signature traits of QD including qudit systems~\cite{OrtheyPRA}, Clifford circuits~\cite{Ferte2024}, skyrmionic and antiferromagnetic systems~\cite{Sotnikov2023} and other many-body models ~\cite{ollivierETR, davide2022, mirkin2021many}. Furthermore, these effects have been demonstrated in several recent experiments~\cite{Unden2019PRL, Ciampini2018PRA, Chen2019SB, chisholm2021witnessing, sq2025}, see Refs.~\cite{duruisseau2023pointer,Doucet2024} for a full classification.

The framework of QD posits that the emergence of objectivity is a consequence of the interaction between the system and the surrounding environment ($E$), assumed to consist of a large number of constituent subsystems. Due to this interaction, information about the system is exchanged with the environment and objectivity emerges when this information is redundantly encoded in the environmental constituents. This results in observers who, by measuring distinct fragments ($F$) of the environment, agree on the state of the system provided that the fragment fractional size $f\!=\!|F|/|E|$ is sufficiently large~\cite{Zurek2009NP}, where $|\cdot|$ denotes the number of constituent parts.

The main tools to characterise the emergence of objectivity in quantum systems involve either quantum information-theoretic measures, such as the (quantum or classical) mutual information~\cite{Zurek_2025, Ollivier2004PRL, Zurek2009NP, touil2022,touil2023black, LeCastro, dianaPRA, chisholm2023meaning}, or by examining the geometric properties of the overall system-environment state~\cite{horodecki2015quantum, korbicz2014objectivity, mironowicz2017monitoring, Korbicz2021roadstoobjectivity, touil2022branching}. In this work, we take an alternative operational approach and recast the problem in the language of quantum metrology. We consider a simple, but generic model~\cite{Campbell2019PRA, touil2022}, that supports the conditions necessary for classical objectivity as set by the QD framework. We focus on the {\it accessibility} of the relevant system information directly, rather than the build up of correlations, in a similar spirit to how consensus~\cite{chisholm2023meaning} focuses on the retrievability of information by the observers.  

Specifically, we examine how observers can access relevant information about the system by measuring the environmental fragment to spewhich they have access. To do this, we consider the scenario typical in quantum sensing~\cite{paris2009quantum,giovannetti2011,Haase2016,Pezze2018,liu2020quantum} where information about a certain parameter $\theta$ has been encoded into the state of the system. If system and environment are initially uncorrelated the initial state reads
\begin{equation}
\rho_0(\theta) =\rho_0^E \otimes (e^{i \theta V_S} \rho_0^S  e^{-i \theta V_S}).
\end{equation}
At time $t$, the initial state evolves through the interaction Hamiltonian $H$ into the state $\rho_t(\theta)=e^{-iHt}\rho_0(\theta) e^{iHt}$. 

In QD, objectivity implies that different observers, each measuring distinct fragments of the environment, agree on the properties of the system. In this case, inferring properties of the state of the system implies accurately determining the value of the unknown parameter $\theta$ from measurements of the environmental fragment state
\begin{eqnarray}
\rho_t^F(\theta)= \Tr_{E/F} \Tr_S \rho_t(\theta), \label{eq:genstate}
\end{eqnarray}
where $t$ is the interaction time and ``$\Tr_x$'' denotes the partial trace over subsystem ``$x$''.  High precision corresponds to a small variance of this estimate from the true value. The quantum Fisher information (QFI), $\mathcal{F}_\theta$, provides a convenient means to lower bound this variance using the Cram\'{e}r-Rao bound~\cite{paris2009quantum}
\begin{equation}
{\rm Var}(\theta) \geq \mathcal{F}_\theta^{-1}.
\end{equation}
The QFI depends on the state $\rho_t^F(\theta)$ and its derivative which can be expressed as
\begin{eqnarray}
\partial_\theta \rho_t^F(\theta) = \Tr_{E/F} \Tr_S \left\{e^{- i H t} \left(\rho_0^E \otimes i [V_S,e^{i \theta V_S} \rho_0 ^S  e^{-i \theta V_S}]  \right) e^{i H t} \right\}. \nonumber \\ \label{eq:derivative}
\end{eqnarray}
The QFI thus provides a means of determining how precisely an observer can determine the state of the system from specific measurements of a given environmental fragment. It can be computed using~\cite{Safranek2018}
\begin{eqnarray}
\mathcal{F}_\theta = 2 {\mathcal V}\left[\partial_\theta \rho_t^F (\theta)\right]^T \cdot \left[\rho_t^F (\theta)^T \otimes \mathbb{I}+\mathbb{I}\otimes \rho_t^F (\theta) \right]^{-1} \cdot {\mathcal V}\left[\partial_\theta \rho_t^F (\theta)\right], \nonumber \\
\end{eqnarray}
where ${\mathcal V}[\cdot]$  denotes vectorization of a matrix and $\mathbb{I}$ is the identity matrix.

The above framework is general and allows us to explore how the interaction time, $t$, and fragment fraction size, $f$, affect the measurement precision for any given interaction Hamiltonian. Furthermore, our framework also to some degree circumvents the high practical and computational resources required for experimental witnesses of objectivity~\cite{Ciampini2018PRA, Unden2019PRL, Chen2019SB, chisholm2021witnessing}, that have so far limited tests for large systems by leveraging known techniques for quantum Fisher information retrieval~\cite{Fisher1,Fisher2,Fisher3}.

We note that the QFI has been previously examined in several related contexts. Broadcasting of information is a key ingredient of QD. In Ref.~\cite{BroadcastingPRA} the conditions for broadcasting QFI (mapping the parameter dependent state to a bipartite state with reduced states of equal QFI to the original state) are established. Similarly on the topic of information retrieval from an evironment, a related quantity termed quantum fisher information kernel is defined \cite{Korbicz2024}. This quantity connects the decoherence on the system to the information accumulated in the environment.

\begin{figure}[t]
    \centering
    \includegraphics[width=0.4\linewidth]{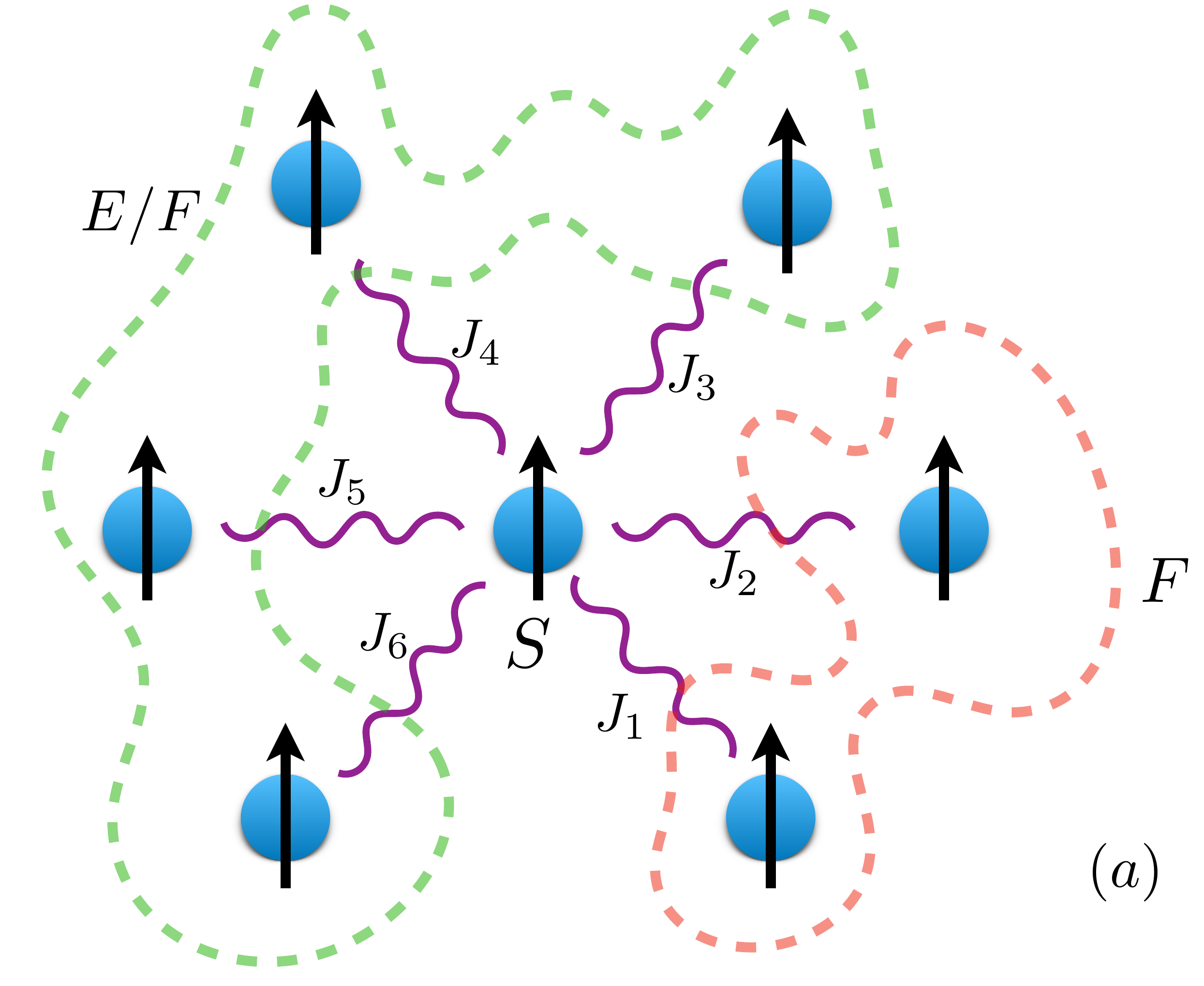}
    \includegraphics[width=0.5\linewidth]{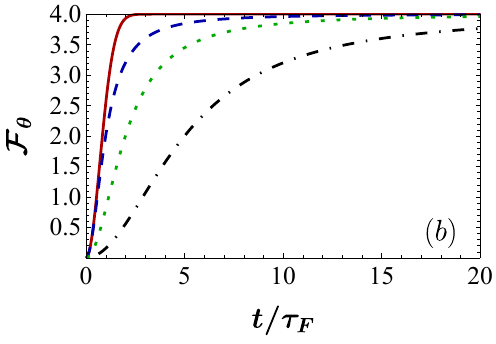}
    \caption{(a) Diagram of the spin-star model for $N=6$ and $f=1/3$. The environmental fragment accessible to an observer is shown in the red region and the remaining environment in the green region. (b) Thermodynamic limit of QFI (red, solid line) against time. Also shown is the corresponding precision for $S_y$ with $\tau_Y/\tau_F=\{1,2,5\}$ (blue dashed, green dotted and black dot-dashed lines). }
    \label{fig:diagram}
\end{figure}

\section{Model and maximal achievable precision} 
We consider the spin-star model \cite{Campbell2018,Campbell2019PRA,Garcia2020PRR,touil2022} where the system is a single qubit and the environment is a collection of $N\!=\!|E|$ qubits, see Fig.~\ref{fig:diagram}(a). We assume the interaction between the system and the environment to be governed by the Hamiltonian
\begin{eqnarray}
H= \sum_{m=1}^{N} \frac{J_m}{\sqrt{N}} \sigma_z^m \sigma_z^{N+1},
\end{eqnarray}
where $J_m$ are the individual coupling strengths assumed to be randomly drawn from some distribution and are suitably scaled to ensure a well defined thermodynamic limit. We further assume an initial product state $\ket{\psi_0}\! =\!  \bigotimes_{k=1}^{N} \ket{+}_k \otimes \left(\alpha_\theta \ket{\downarrow}_{N+1} + \beta_\theta \ket{\uparrow}_{N+1}\right)$ where $\ket{+}_m= ( \ket{\downarrow}_m+ \ket{\uparrow}_m)/\sqrt{2}$. We choose $V_S\!=\!\sigma_x$ and $\rho_0^S\!=\! \ketbra{\downarrow}{\downarrow}$, such that $\alpha_\theta\!=\!\cos \theta$ and $\beta_\theta\!=\!i \sin \theta$. 

The reduced state of the system is
\begin{eqnarray}
    \rho_t^S = \left( \begin{array}{cc}
 \cos^2 \theta & i \cos \theta \sin \theta \, e^{-\Gamma(t)} \\
  -i \cos \theta \sin \theta \, e^{-\Gamma(t)} & \sin^2 \theta \\
\end{array}  \right),
\label{eq:systemstate}
\end{eqnarray}
expressed in the basis $\ket{\downarrow}\!=\!(\begin{array}{cc}  0 & 1 \end{array})^T$ and $\ket{\uparrow}\!=\!(\begin{array}{cc}  1 & 0 \end{array})^T$ where $\Gamma(t)$ corresponds to the decoherence factor
\begin{equation}
    \Gamma(t) = -  \ln \left[\prod_{k=1}^N \cos\left( J_k t/ \sqrt{N}\right) \right]~~\underset{N\to\infty}{\longrightarrow}~~
 \left(\frac{t}{\tau_D}\right)^2.
 \label{eq:quantumFI}
\end{equation}
In taking the thermodynamic limit we use the approximation $\ln\left( \cos(x)\right)\!\approx\! -x^2/2$. 
This limit allows us to identify a characteristic decoherence timescale $\tau_D\!=\!1/\sqrt{2 \avg{J^2}}$, which is also valid for finite environment sizes provided $t \!\ll\!\! \sqrt{N}$. The maximal precision achievable by measuring an environmental fragment can at most be equal to the QFI of the system state given in Eq.~\eqref{eq:systemstate}, which by direct calculation is a constant, $\mathcal{F}_\text{max}\!=\!4$, for all times. A deviation from maximum redundancy in the context of mutual information, so-called ``information deficit", has been previously explored in \cite{Zwolak2013SR}.

We now determine how accurately an observer with access to a given fragment can determine the state of the system. The reduced state of a fragment of size $|F|$ is given by
\begin{eqnarray}
\rho_t^F = \cos^2(\theta) \bigotimes_{m=1}^{|F|} \Omega_m(t) + \sin^2(\theta) \bigotimes_{m=1}^{|F|} \Omega_m(-t),
\label{eq:fragmentstate}
\end{eqnarray}
where
\begin{eqnarray}
 \Omega_m(t) = \frac{1}{2} \left(
\begin{array}{cc}
 1 & e^{2 i J_m t/ \sqrt{N}} \\
  e^{-2 i
   J_m t/\sqrt{N}} & 1
\end{array}
\right).
\end{eqnarray}
Calculating the QFI for Eq.~\eqref{eq:fragmentstate} gives (see Appendix~\ref{appA}), 
\begin{equation}
    \mathcal{F}_\theta = 4 \left[1-\prod_{k=1}^{|F|} \cos^2(2 J_k t/\sqrt{N}) \right] 
    ~~\underset{N\to\infty}{\longrightarrow}~~ 4\left[1-e^{- (t/\tau_F)^2 }\right],
\end{equation}
where $\tau_F\!=\!1/(2\sqrt{ f \avg{J^2}})$ emerges as the natural QFI timescale for the fragment. Note that for this specific case, the QFI is independent of $\theta$. The increase in QFI is clearly related to decoherence of the system as $\tau_F\!=\!\tau_D/\sqrt{2 f}$. This result is consistent with the fact that system decoherence is necessary, but not sufficient, for objectivity to emerge~\cite{Campbell2019PRA}. While the system will reach an effectively classical state on timescales on the order of $\tau_D$, small fragments ($f<1/2$) require more time before the relevant system information is imprinted on their degrees of freedom i.e. $\tau_F > \tau_D$.

\section{Precision for a specific observable} 
While the QFI bounds the maximal precision, it necessarily assumes that the optimal observable is measured, which for the model under consideration is time dependent, see Appendix~\ref{appA}. However, given that independent observers are free to perform any measurement on their fragments, it is relevant to consider how a specific measurement choice impacts the ability of an observer to determine the state of the system. Therefore, we now consider measurements of a specific observable, $A$, which from the error propagation formula has a variance
\begin{eqnarray}
    \text{Var}\left(\theta\right) = \frac{\text{Var}\left(A\right)}{\left|\partial_\theta \avg{A}\right|^2},
\label{eq:classicalprecision}
\end{eqnarray}
where $\text{Var}\left(A\right)\! =\!\avg{A^2} - \avg{A}^2$. Throughout the text, we will refer to the inverse of Eq.~\eqref{eq:classicalprecision} as precision.

A natural choice is to consider combinations of static local observables of the form $S_k\!=\!\sum_{i=1}^{|F|} \sigma_i^k$ for $k\!=\!\{x,y,z\}$. For $k\!=\!x$ or $z$ we find that the corresponding expected value has no $\theta$ dependence and, therefore, an observer with access to only these measurements can attain no information regarding the state of the system. In contrast, the choice of $k\!=\!y$ corresponds to the optimal static local observable in the thermodynamic limit, see Appendix~\ref{appB}. For finite sizes, the expectation value can be explicitly calculated
\begin{eqnarray}
     \avg{S_y}= -\cos(2 \theta) \sum_{m=1}^{|F|} \sin\left( 2 J_m t /\sqrt{N} \right),
\end{eqnarray}
and, through Eq.~\eqref{eq:classicalprecision}, we can determine the associated variance, see Appendix~\ref{appB},
\begin{eqnarray}
   \text{Var}\left(\theta\right) &=&  \frac{1}{4} \left[ 1+\frac{|F|-\sum_n  \sin^2\left( 2 J_n t /\sqrt{N} \right)}{\sin^2(2 \theta)\left| \sum_n  \sin\left( 2 J_n t/ \sqrt{N}  \right) \right|^2} \right] \nonumber \\
   &\underset{N\to\infty}{\longrightarrow}& \frac{1}{4} \left[ 1+\frac{1}{ (t/\tau_Y)^2} \right].
   \label{eq:classicalFI}
\end{eqnarray}
where we have assumed $\avg{J}\!\neq\! 0$~\footnote{We remark that this assumption is not critical. However, for a zero mean distribution of coupling strengths, the expression for the thermodynamic limit differs from the one presented in the main text.}. Similar to the case of optimal measurements, we find that  $\tau_Y\!\!=\!\!\left(2 |\sin(2 \theta)\avg{J}|\sqrt{f}\right)^{-1}$. 

This captures the relevant timescale over which system information is imprinted on the fragment. Note that $\tau_Y$ diverges for $f=0$ (since there is no fragment to measure) and $\theta=\{0,\pi/2\}$ since the $\theta$ derivative of the expected value vanishes. We also note that, since $\avg{J^2}\!\geq\! \avg{J}^2$, it follows that $\tau_F\!\leq\! \tau_Y$, i.e. the optimal measurements attain information at a faster rate. In fact the QFI scales exponentially, while the precision of the suboptimal observable, $S_y$, scales algebraically, as shown in Fig.~\ref{fig:diagram}(b). Thus, the timescale for the QFI to reach its maximum value is always shorter than the corresponding timescale for maximising the precision of measuring $S_y$. This implies that if observers can only employ static local observables to infer the system's information, they will need to wait for longer times. Importantly, we see that even suboptimal measurements will still achieve the maximum precision given sufficiently long times in the thermodynamic limit. This provides evidence that the emergence of objectivity is indeed a generic feature.

\section{Numerical results} 
With the above framework in hand, we now examine how accurately the asymptotic results hold for finite sizes. We consider a Gaussian distribution of coupling strengths with mean $\avg{J}\!=\!\mathcal{J}$ and variance $\avg{J^2}-\avg{J}^2\!=\!\sigma^2$. It follows from Eqs.~\eqref{eq:quantumFI} and \eqref{eq:classicalFI} that both quantities exhibit initial quadratic growth in the thermodynamic limit
\begin{equation}
   \mathcal{F}_\theta \approx 4 (t/\tau_F)^2, \qquad
    \text{Var}\left(\theta\right) ^{-1} \approx 4 (t/\tau_Y)^2.
\end{equation}

\begin{figure}[b]
    \centering
    \includegraphics[width=0.49\columnwidth]{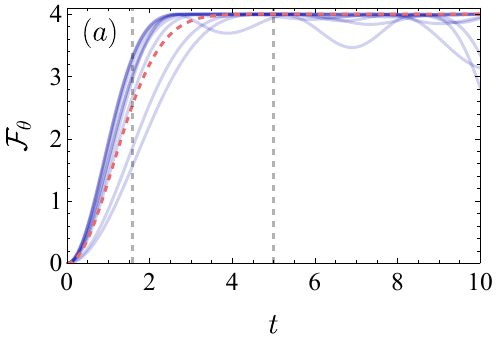}
     \includegraphics[width=0.49\columnwidth]{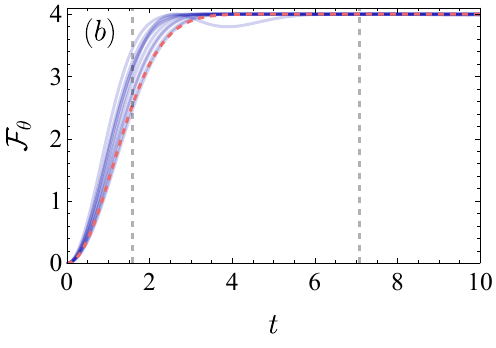}\\
      \includegraphics[width=0.49\columnwidth]{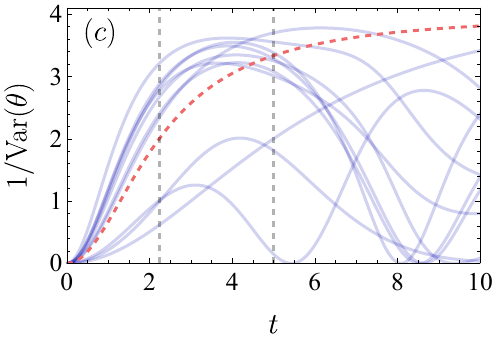} 
      \includegraphics[width=0.49\columnwidth]{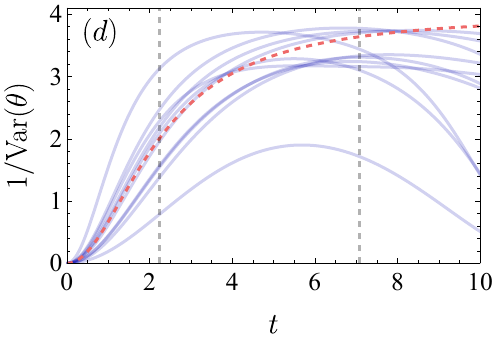}
    \caption{QFI, $\mathcal{F}_\theta$, and precision, $1/\text{Var}\left(\theta\right)$, against time $t$ for $10$ different realisations of the coupling strengths $J$ (blue semitransparent lines). The asymptotic behaviour, i.e. thermodynamic limit, is also shown by the red, dashed line. We choose $f=0.2$, $\mathcal{J}=0.5$ and $\sigma=0.5$. Different columns corresponds to increasing system size $N=25$ (left) and $N=50$ (right). Dashed vertical grey lines at $t=\tau_{F,Y}$ and $t=\sqrt{N}$.}
    \label{fig:figure1}
\end{figure}

In Fig.~\ref{fig:figure1}(a,b) we show the dynamics of the QFI for a fixed fragment size, $f\!=\!0.2$. The red dashed line corresponds to the thermodynamic limit while the solid, blue semitransparent curves correspond to 10 random choices of the coupling strengths drawn from a Gaussian distribution and the vertical dotted lines are at $\tau_F$ and $\sqrt{N}$, respectively. For a small environment, $N\!=\!25$ shown in panel (a), we see that while the initial dynamics ($t<\tau_F$) of the QFI exhibits a similar quadratic growth to the thermodynamic limit, the finite size effects quickly become apparent. For the single realisations the QFI transiently reaches its maximum value, typically within the expected time window, i.e. $\tau_F \!<\! t \!<\! \sqrt{N}$. This implies that the ability of observers to accurately determine and agree on the state of the system is delicately dependent on the precise time at which the fragments are measured. However, even a modest increase in the system size is sufficient for the finite size environments to be well approximated by the $N\!\to\!\infty$ limit. For $N\!=\!50$, we find the QFI for the single realisations typically saturates, only showing some small oscillations when $t\!>\! \sqrt{N}$. This behavior becomes more stable as $N$ is increased.
\begin{figure}[t]
    \centering
    \includegraphics[width=0.49\linewidth]{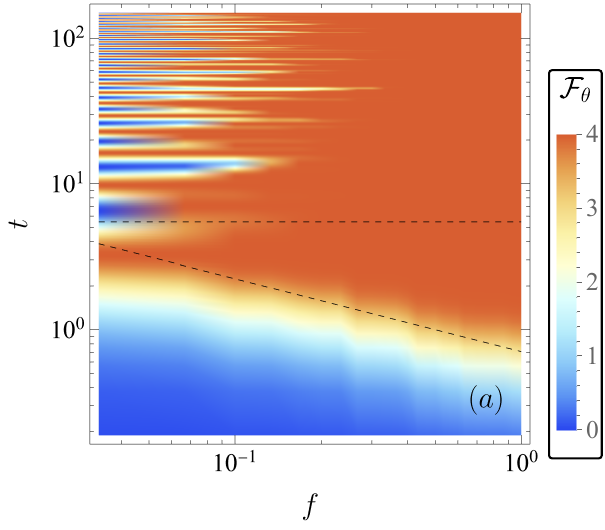}
    \includegraphics[width=0.49\linewidth]{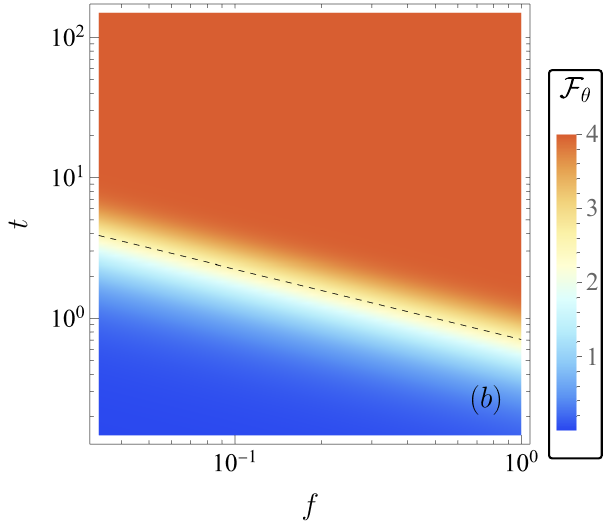}
    \caption{QFI against fragment fraction $f$ and time $t$ for (a) a single realizations of coupling strengths for $\mathcal{J}=0.5$, $\sigma=0.5$ and $N=30$ (b) thermodynamic limit. Black dashed lines show $t=\tau_F$ and $t=\sqrt{N}$.}
    \label{fig:figure2}
\end{figure}

For the case of the optimal static local measurement, $S_y$, we find a similar trend with increasing environment size, shown in Fig.~\ref{fig:figure1}(c,d). However, a notable difference for this choice of measurement is the clearer onset of finite size effects. This is particularly evident in panel (d) where we see that the asymptotic result well approximates the single realisations for an environment of size $N\!=\!50$, while for $t\!\sim\!\sqrt{N}$ the results rapidly start to deviate.

Figure~\ref{fig:figure2} shows the QFI as a function of both time and fragment fraction, $f$. For a finite environment we show a representative simulation with $N\!=\!30$, again drawing the couplings from a Gaussian distribution. We clearly see two regimes present. The first is delineated by the diagonal dashed line which corresponds to $t\!=\!\tau_F(f)$ and shows that observers with access to a given fragment size will only achieve the optimal precision after a sufficient time has passed. The horizontal dashed line at $t\!=\!\sqrt{N}$ shows that beyond this timescale finite size effects will impact the precision with which observers can determine the state of the system. Moving to the thermodynamic limit, panel (b) clearly shows that $\tau_F$ defines the characteristic timescale for the emergence of classical objectivity.

\begin{figure}[t]
    \centering
    \includegraphics[width=0.85\linewidth]{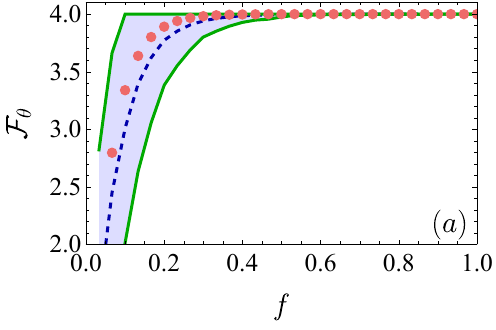}\\
    \includegraphics[width=0.85\linewidth]{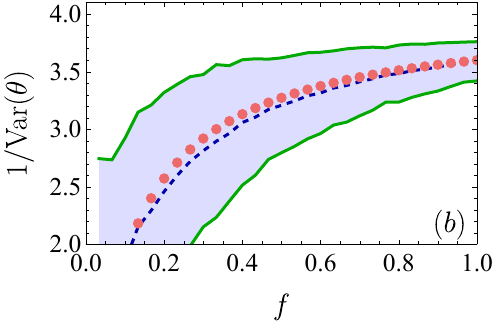}
    \caption{The behaviour of (a) QFI and (b) precision with $S_y$ measurement with increasing fragment fraction $f$ at a late time $t=3$ with $N=30$,  $\mathcal{J}=0.5$ and $\sigma=0.5$.  Dashed blue line shows the average behaviour from  $2000$ random realisations of the coupling. Realisations within one standard deviation of the mean reside in the blue shaded area bordered with green lines and red dots show the thermodynamic limit.}
    \label{fig:figure3}
\end{figure}

The hallmark of quantum Darwinism is the existence of a so called redundancy plateau of (mutual) information i.e. that the information about the system shared with the environment which quickly saturates~\cite{Zurek2009NP}. In Fig. \ref{fig:figure3} we show that the QFI and precision exhibit the same characteristic behavior. The interaction timescale $t$ is chosen such that $\tau_{F,Y} < t < \sqrt{N}$. In panel (a) we see that for an environment of $N=30$ qubits, the QFI quickly grows with fragment size, saturating to the maximum at $f\!\approx\!0.3$. The effect is less pronounced for the precision as shown in panel (b). This panel also demonstrates that small environmental sizes limits the maximal achievable precision. Together these results indicate that QD is a generic feature. For small environments, the maximal precision can be attained however, this requires observers to measure the optimal observable. For the arguably more relevant scenario of mesoscopic environments, while optimal observables will saturate to the maximal achievable precision quicker, even suboptimal measurements performed by observers on small fragments of the environment will be sufficient to attain the same level of precision. We remark that these results do not depend on specific interaction times, the only requirement is that the interaction time exceeds the relevant characteristic timescales dictated by the measurement choice. Note that any parameter independent noise channel on the environmental fragment will generically reduce the QFI.

\section{Conclusions} 
We have shown how the emergence of classicality can be more precisely characterized using tools from quantum metrology. In particular, we have established that the quantum Fisher information provides a clear operational meaning for how accurately the observer can infer the state of the system based on the measurements they perform on the environmental fragments. Our approach is a first step in the integration of quantum objectivity with quantum metrology, therefore opening new avenues for employing novel tools in the study of the emergence of classicality. In particular, it provides a framework to precisely characterize consensus between multiple observers performing different measurements on their fragment \cite{chisholm2023meaning, touil2025consensus}.

We have focused on the case of estimating a single parameter in the setting where the randomized system-environment interactions dominate over any intra-environmental couplings. Future work will focus on extending this to models with complex environmental structures with intra-environmental interactions \cite{riedel2012,mihailescu2024quantum}, as well as studying the objectivity of multiparameter estimation using the quantum Fisher information matrix \cite{paris2009quantum,liu2020quantum,mihailescu2026}. Moving away from exactly solvable models will require some concession. More complex baths could be modeled using open system theory techniques, requiring certain approximations. Alternatively, arbitrary baths could be considered numerically, via exact diagonalisation (see e.g. Appendix~\ref{appC}), but this approach is restricted to relatively small system sizes.

\section*{Acknowledgments}
This work was supported by the John Templeton Foundation under Grant Nos. 62422 and 63626. DAC acknowledges support from  the ``Italian National Quantum Science and Technology Institute (NQSTI)" (PE0000023) - SPOKE 2 through project ASpEQCt, and from Taighde \'Eireann - Research Ireland under grant number GOIPD/2025/1353. A.T. is supported by the U.S DOE under the LDRD program at Los Alamos. This work was supported by the U.S. Department of Energy, Office of Basic Energy Sciences, Quantum Information Science program in Chemical Sciences, Geosciences, and Biosciences, under Award No. DE-SC0025997.

\appendix

\section{Proof of QFI expression }\label{appA}

The relevant state of the fragment can be expressed as $\rho^F_t=p_\theta \ketbra{\phi_t}{\phi_t}+(1-p_\theta)\ketbra{\phi_{-t}}{\phi_{-t}}$ where $p_\theta=\cos^2 \theta$ and $\ket{\phi_t}= \bigotimes_{m=1}^{|F|} e^{ i J_m \sigma_z^m t/\sqrt{N}} \ket{+}_m$. This is clearly a rank two matrix since it is a mixture of two pure initial states.  Let us use an orthonormal basis $\{\ket{\phi_t}, \ket{\phi_t^\perp}\}$.  If we define the overlap $c(t)=\braket{\phi_t}{\phi_{-t}}$, we can write
\begin{eqnarray}
\ket{\phi_{-t}} = c(t) \ket{\phi_t} + \sqrt{1-\left|c(t) \right|^2} \ket{\phi_t^\perp}.
\end{eqnarray}
This overlap can be explicitly computed as
\begin{equation}
c(t) = \prod_{m=1}^{|F|} \bra{+} e^{-2  i J_m \sigma_z^m t/\sqrt{N}}  \ket{+} = \prod_{m=1}^{|F|} \cos(2 J_m t/\sqrt{N} ).
\end{equation}
Working in this basis, the density matrix is
\begin{eqnarray}
\rho^F_t &=& \left[p_\theta+(1-p_\theta) |c(t)|^2 \right] \ketbra{\phi_t}{\phi_t} + \left[1-|c(t)|^2\right](1-p_\theta) \ketbra{\phi_t^\perp}{\phi_t^\perp} \nonumber \\ &+& (1-p_\theta) c(t) \sqrt{1-|c(t)|^2} \ketbra{\phi_t}{\phi_t^\perp} \nonumber \\
&+& (1-p_\theta) c^*(t) \sqrt{1-|c(t)|^2} \ketbra{\phi_t^\perp}{\phi_t}.
\end{eqnarray}
% Using the shorthand $\rho=\rho_t^F (\theta)$ and $d \rho =\partial_\theta \rho_t^F (\theta)$,  the QFI can be then directly computed using \cite{Safranek2018}
% \begin{eqnarray}
% \mathcal{F}_\theta = 2 (\vec{d\rho})^T \cdot \left[\rho^T \otimes \mathbb{I}+\mathbb{I}\otimes \rho \right]^{-1} \cdot \vec{d\rho},
% \end{eqnarray}
% where the arrows indicate a vectorised matrix. 
The QFI can be computed directly using \cite{Safranek2018}
\begin{eqnarray}
\mathcal{F}_\theta = 2 {\mathcal V}\left[\partial_\theta \rho_t^F (\theta)\right]^T \cdot \left[\rho_t^F (\theta)^T \otimes \mathbb{I}+\mathbb{I}\otimes \rho_t^F (\theta) \right]^{-1} \cdot {\mathcal V}\left[\partial_\theta \rho_t^F (\theta)\right], \nonumber \\
\end{eqnarray}
where ${\mathcal V}[\cdot]$  denotes vectorization of a matrix. This gives
\begin{eqnarray}
\mathcal{F}_\theta &=& \frac{\dot{p_\theta}^2}{p_\theta^2-p_\theta} \left[c(t)^2-1\right] \\
&=& 4 [1-c(t)^2],
\end{eqnarray}
which is the result stated in the main text.  Note that this is independent of $\theta$. To obtain the thermodynamic limit, we rewrite this as
\begin{eqnarray}
\mathcal{F}_\theta &=& 4 \left[1-\prod_{k=1}^{|F|} \cos^2(2 J_k t/\sqrt{N}) \right]  \\
&=& 4 \left\{1-\exp\left[\sum_{k=1}^{|F|} \ln \cos^2(2 J_k t/\sqrt{N})\right]\right\} \\
&\rightarrow & 4 \left[1-e^{-4 f t^2 \avg{J^2}}\right].
\end{eqnarray}

Using a similar formula,  one can identify the symmetric logarithmic derivative as
\begin{eqnarray}
L_\theta &=& 2[c(t)^2-1] \tan\theta \ketbra{\phi_t}{\phi_t} \nonumber \\ &+& \left[1-c(t)^2+(1+c(t)^2) \cos(2 \theta) \right]\csc \theta \sec \theta \ketbra{\phi_t^\perp}{\phi_t^\perp} \nonumber \\ &+& 2 c(t) \sqrt{1-c(t)^2} \tan \theta \left[\ketbra{\phi_t}{\phi_t^\perp} + \ketbra{\phi_t^\perp}{\phi_t} \right].
\end{eqnarray}
The optimal observable is then given by 
\begin{eqnarray}
 X_\theta = \theta  + \frac{L_\theta}{\mathcal{F}_\theta}.
\end{eqnarray}
In the thermodynamic and long time limit $c(t)^2$ vanishes, reducing this to
\begin{eqnarray}
X_\theta \rightarrow \left(\theta - \frac{\tan\theta}{2}\right) \ketbra{\phi_t}{\phi_t} +\left( \theta+\frac{ \cot \theta}{2} \right) \ketbra{\phi_{-t}}{\phi_{-t}}.
\end{eqnarray}

\section{Precision for a local observable\label{appB}}

To start with, let us define the local expectation values $\avg{\sigma^p_m(t)} \equiv \Tr \left[\sigma^p_m \Omega_m(t)\right]$ which can be explicitly computed as
\begin{eqnarray}
     \avg{\sigma^x_m(t)} &=& \avg{\sigma^x_m(-t)} =\cos(2 J_m t/\sqrt{N}), \\
     \avg{\sigma^y_m(t)} &=& -\avg{\sigma^y_m(-t)}= -\sin(2 J_m t/\sqrt{N}), \\
     \avg{\sigma^z_m(t)} &=& 0.
\end{eqnarray}
We construct our local observable to be $A_q = \sum_{i=1}^{|F|} \left[ q \sigma_i^x +(1-q) \sigma_i^y \right]$. The average of this observable is then
\begin{widetext}
\begin{eqnarray}
    \avg{A_q} &=& \Tr \left[ A_q \, \rho_t^F(\theta)\right] \\
     &=& \cos^2\theta \sum_{m=1}^{|F|} \left[q \avg{\sigma^x_m(t)} +(1-q)\avg{\sigma^y_m(t)}\right]  + \sin^2 \theta \sum_{m=1}^{|F|} \left[q \avg{\sigma^x_m(-t)} +(1-q)\avg{\sigma^y_m(-t)}\right] \\
    &=& q \sum_{m=1}^{|F|}\avg{\sigma^x_m(t)} +(1-q) \cos(2 \theta) \sum_{m=1}^{|F|}\avg{\sigma^y_m(t)} .
\end{eqnarray}
This would appear to indicate that the optimal observable is $q=0$, since this part has the $\theta$ dependent term. However, we must also consider fluctuations in this observable.

To this end, the second moment can be calculated as
\begin{equation}
         \avg{A_q^2} = \Tr \left[ A_q^2 \, \rho_t^R(\theta)\right]
     = q^2 \Tr\left[\sum_{i,j} \sigma_i^x \sigma_j^x \rho_t^R(\theta)\right]+ (1-q)^2 \Tr\left[\sum_{i,j} \sigma_i^y \sigma_j^y \rho_t^R(\theta)\right] +q(1-q) \Tr\left[\sum_{i,j} \left\{\sigma_i^x \sigma_j^y+ \sigma_i^y\sigma_j^x\right\} \rho_t^R(\theta)\right].
\end{equation}
Let us compute each part separately. The first part is
\begin{eqnarray}
   \Tr\left[\sum_{i,j} \sigma_i^x \sigma_j^x \rho_t^R(\theta)\right] &=& |F| + \cos^2(\theta)\sum_{i\neq j}^{|F|} \avg{\sigma^x_i(t)} \avg{\sigma^x_j(t)} + \sin^2(\theta)\sum_{i\neq j}^{|F|} \avg{\sigma^x_i(-t)} \avg{\sigma^x_j(-t)} \\
   &=& |F| + \sum_{i\neq j}^{|F|} \avg{\sigma^x_i(t)} \avg{\sigma^x_j(t)}.
\end{eqnarray}
The next term is
\begin{eqnarray}
   \Tr\left[\sum_{i,j} \sigma_i^y \sigma_j^y \rho_t^R(\theta)\right] &=& |F| + \cos^2(\theta)\sum_{i\neq j}^{|F|} \avg{\sigma^y_i(t)} \avg{\sigma^y_j(t)} + \sin^2(\theta)\sum_{i\neq j}^{|F|} \avg{\sigma^y_i(-t)} \avg{\sigma^y_j(-t)} \\
   &=& |F| + \sum_{i\neq j}^{|F|} \avg{\sigma^y_i(t)} \avg{\sigma^y_j(t)} .
\end{eqnarray}
The final term is 
\begin{eqnarray}
    \Tr\left[\sum_{i,j} \left\{\sigma_i^x \sigma_j^y+ \sigma_i^y\sigma_j^x\right\} \rho_t^R(\theta)\right] &=& \cos^2(\theta) \sum_{i \neq j} \left\{\avg{\sigma^x_i(t)} \avg{\sigma^y_j(t)} + \avg{\sigma^y_i(t)} \avg{\sigma^x_j(t)}\right\} + \sin^2(\theta) \sum_{i \neq j} \left\{\avg{\sigma^x_i(-t)} \avg{\sigma^y_j(-t)} + \avg{\sigma^y_i(-t)} \avg{\sigma^x_j(-t)}\right\} \nonumber \\
    &=& 2\cos(2\theta) \sum_{i \neq j} \avg{\sigma^x_i(t)} \avg{\sigma^y_j(t)}.
\end{eqnarray}
All together then we get
\begin{eqnarray}
    \avg{A_q^2} &=& |F| + q^2 \sum_{i\neq j}^{|F|} \avg{\sigma^x_i(t)} \avg{\sigma^x_j(t)} +(1-q)^2 \sum_{i\neq j}^{|F|} \avg{\sigma^y_i(t)} \avg{\sigma^y_j(t)} + 2 q(1-q) \cos(2 \theta) \sum_{i\neq j}^{|F|} \avg{\sigma^x_i(t)} \avg{\sigma^y_j(t)}, \\
    \avg{A_q}^2 &=& q^2 \left( \sum_{m=1}^{|F|}\avg{\sigma^x_m(t)} \right)^2 +(1-q)^2 \cos^2(2 \theta)\left( \sum_{m=1}^{|F|}\avg{\sigma^y_m(t)} \right)^2 + 2q(1-q) \cos(2\theta) \left( \sum_{i=1}^{|F|}\avg{\sigma^x_i(t)} \right) \left( \sum_{j=1}^{|F|}\avg{\sigma^y_j(t)} \right).
\end{eqnarray}
Finally the variance is given by
\begin{eqnarray}
    \text{Var}\left(\theta\right) 
   &=& \frac{\text{Var}\left(A_q\right)}{\left|\partial_\theta \avg{A_q}\right|^2} \\
   &=& \frac{|F|-q^2 \sum_{m=1}^{|F|} \avg{\sigma^x_m(t)}^2+(1-q)^2\left[\sin^2(2\theta)\left|\sum_{m=1}^{|F|} \avg{\sigma^y_m(t)}\right|^2-\sum_{m=1}^{|F|} \avg{\sigma^y_m(t)}^2\right]-2q(1-q) \cos(2 \theta) \sum_{m=1}^{|F|} \avg{\sigma^x_m(t)} \avg{\sigma^y_m(t)} }{4 \sin^2(2 \theta)(1-q)^2 \left|\sum_{m=1}^{|F|} \avg{\sigma^y_m(t)}\right|^2} \nonumber \\
   &=& \frac{1}{4} \left[1+\frac{|F|-q^2 \sum_{m=1}^{|F|} \avg{\sigma^x_m(t)}^2-(1-q)^2\sum_{m=1}^{|F|} \avg{\sigma^y_m(t)}^2-2q(1-q) \cos(2 \theta) \sum_{m=1}^{|F|} \avg{\sigma^x_m(t)} \avg{\sigma^y_m(t)} }{ \sin^2(2 \theta)(1-q)^2 \left|\sum_{m=1}^{|F|} \avg{\sigma^y_m(t)}\right|^2}\right] \\
   &=& \frac{1}{4} \left\{1+\frac{\sum_{m=1}^{|F|}\left[1-q^2 +(2q-1)\avg{\sigma^y_m(t)}^2 -2q(1-q) \cos(2 \theta) \avg{\sigma^x_m(t)} \avg{\sigma^y_m(t)} \right]      }{ \sin^2(2 \theta)(1-q)^2 \left|\sum_{m=1}^{|F|} \avg{\sigma^y_m(t)}\right|^2}\right\}.
\end{eqnarray}
The thermodynamic limit of this can be derived using the approximation $\sin x \approx x$ and discarding terms that vanish in the limit $N\rightarrow \infty$. Note that we assume a non-zero mean for the distribution of couplings $J$. In our case then this reduces to,
\begin{eqnarray}
    \text{Var}\left(\theta\right) &\underset{N\to\infty}{\longrightarrow}& \frac{1}{4} \left[1+ \frac{1}{4 t^2 \sin^2(2 \theta) \avg{J}^2 f} \frac{1+q}{1-q} \right].
\end{eqnarray}
We can clearly see that the optimal choice in the thermodynamic limit is $q=0$. 
\end{widetext}

\section{Effect of interactions between environmental fragments\label{appC}}
\begin{figure}[h!]
    \centering
    \includegraphics[width=0.75\linewidth]{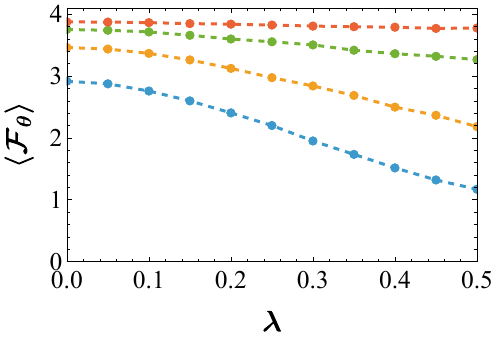}
    \caption{Average QFI over $500$ realisations against intra-environmental interaction strength $\lambda$ for different fraction sizes $N f=\{2,3,4,5\}$(blue, orange, green and red). Other parameters: $N=6$, $t_f=5$ and $\mu=0.2$.}
    \label{fig:APP}
\end{figure}

Here we will briefly consider the case of weak interactions between the environmental qubits. Our modified Hamiltonian is then,
\begin{eqnarray}
    H= \sum_{m=1}^{N} \frac{J_m}{\sqrt{N}} \sigma_z^m \sigma_z^{N+1} + \lambda\sum_{l<m<N+1} \frac{\chi_{l,m}}{\sqrt{N}}\sigma_z^l \sigma_z^m,
\end{eqnarray}
where $\chi_{l,m}$ are independent Gaussian random variables with a variance $\mu^2$. In Fig. \ref{fig:APP}, see how the average QFI behaviour changes when the Hamiltonian deviates from the idealised case described in the main text. For a large perturbation $\lambda$, the scaling of the QFI with fragment size is worse. Nevertheless, small perturbations do not drastically behaviour of the QFI.
   
\begin{figure}[h!]
    \centering
    \includegraphics[width=\linewidth]{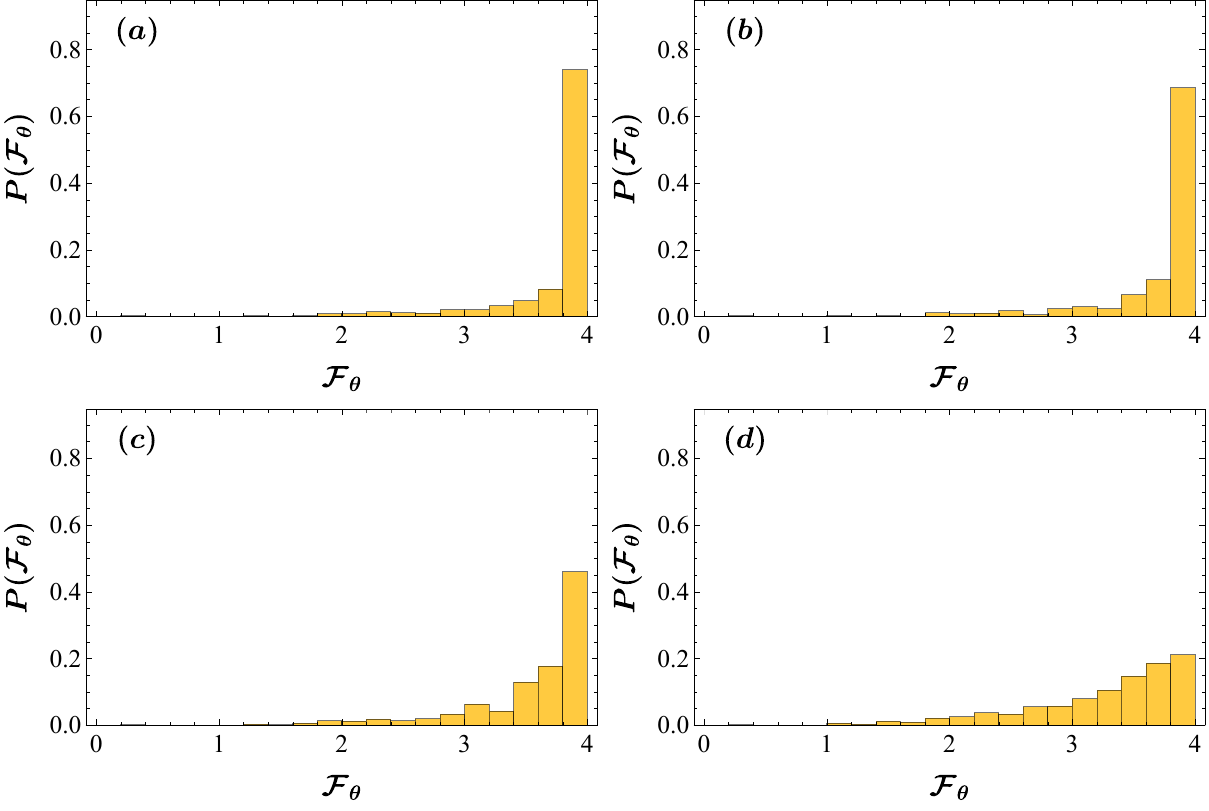}
    \caption{Reconstructed probability distribution of QFI from 500 realisations. Each panel shows a different perturbation strength $\lambda=0,0.1, 0.25,0.5$.  Other parameters are $N f=4$, $N=6$, $t_f=5$ and $\mu=0.2$.}
    \label{fig:APP2}
\end{figure}

In Fig. \ref{fig:APP2}, we can see the distribution of QFI for randomized couplings and how this varies as the perturbation strength is increased. As expected, the clustering around the maximum value of $4$ reduces as the perturbation strength increases.

\clearpage

\let\oldaddcontentsline\addcontentsline     % Store \addcontentsline 
\renewcommand{\addcontentsline}[3]{}     

\bibliography{references_qfidarwinism}

\end{document}